# A Simple Formula for Generating Chern Characters by Repeated Exterior Differentiation

C. C. Briggs
*Center for Academic Computing, Penn State University, University Park, PA 16802*

**Abstract.** A simple formula is given for generating Chern characters by repeated exterior differentiation for *n*-dimensional differentiable manifolds having a general linear connection.

This paper presents a simple formula for generating Chern characters by repeated exterior differentiation for *n*-dimensional differentiable manifolds having a general linear connection.

The Chern characters $ch_{(p)}$ of such a manifold $M$ are $(2p)$-forms as defined, for $p > 0$, by[1-3]

$$\begin{aligned} ch_{(p)} &= \frac{i^p}{2^p \pi^p p!} \operatorname{tr}(\Omega^p) \\ &= \frac{i^p}{2^p \pi^p p!} \Omega_{i_p}{}^{i_1} \wedge \Omega_{i_1}{}^{i_2} \wedge \Omega_{i_2}{}^{i_3} \wedge \ldots \wedge \Omega_{i_{p-3}}{}^{i_{p-2}} \wedge \Omega_{i_{p-2}}{}^{i_{p-1}} \wedge \Omega_{i_{p-1}}{}^{i_p} \\ &= \frac{i^p}{2^p \pi^p p!} \Omega_{i_1}{}^{i_2} \wedge \Omega_{i_2}{}^{i_3} \wedge \ldots \wedge \Omega_{i_{p-3}}{}^{i_{p-2}} \wedge \Omega_{i_{p-2}}{}^{i_{p-1}} \wedge \Omega_{i_{p-1}}{}^{i_p} \wedge \Omega_{i_p}{}^{i_1}, \end{aligned} \tag{1}$$

where $\Omega_a{}^b$ is the curvature 2-form of $M$.

The 1$^{st}$ ordinary exterior differentials of the basis tangent vectors $\mathsf{e}_a$ of $M$ are given by[4-6]

$$\mathrm{d}\, \mathsf{e}_a = \mathsf{e}_b\, \omega_a{}^b, \tag{2}$$

the contractions of which with the basis 1-forms $\omega^b$ of $M$ are given by

$$\langle \omega^b, \mathrm{d}\, \mathsf{e}_a \rangle = \omega_a{}^b \tag{3}$$

and in view of which the 1$^{st}$ absolute exterior differentials of $\mathsf{e}_a$ are given by

$$\begin{aligned} \mathrm{D}\, \mathsf{e}_a &= \mathrm{d}\, \mathsf{e}_a - \mathsf{e}_b\, \omega_a{}^b \\ &= 0, \end{aligned} \tag{4}$$

where $\omega_a{}^b$ is the connection 1-form of $M$ and where the contractions of $\omega^b$ with $\mathsf{e}_a$ are given by

$$\langle \omega^b, \mathsf{e}_a \rangle = \delta^b_a, \tag{5}$$

where $\delta^b_a$ is the Kronecker delta.

The 2$^{nd}$ ordinary exterior differentials of $\mathsf{e}_a$ are given by[7-16]

$$\begin{aligned} \mathrm{d}^2\, \mathsf{e}_a &= \mathrm{d}\, \mathrm{d}\, \mathsf{e}_a \\ &= \mathrm{d}\, \mathsf{e}_b\, \omega_a{}^b \\ &= (\mathrm{d}\, \mathsf{e}_b) \wedge \omega_a{}^b + \mathsf{e}_b\, \mathrm{d}\, \omega_a{}^b \\ &= \mathsf{e}_c\, \omega_b{}^c \wedge \omega_a{}^b + \mathsf{e}_b\, \mathrm{d}\, \omega_a{}^b \\ &= \mathsf{e}_b\, (\mathrm{d}\, \omega_a{}^b + \omega_c{}^b \wedge \omega_a{}^c) \\ &= \mathsf{e}_b\, \Omega_a{}^b, \end{aligned} \tag{6}$$

the contractions of which with $\omega^b$ are given by

$$\langle \omega^b, \mathrm{d}^2\, \mathsf{e}_a \rangle = \Omega_a{}^b \tag{7}$$

and where

$$\begin{aligned} \Omega_a{}^b &= \mathrm{d}\, \omega_a{}^b + \omega_c{}^b \wedge \omega_a{}^c \\ &= \mathrm{D}\, \omega_a{}^b + \omega_a{}^c \wedge \omega_c{}^b \\ &= \mathrm{d}\, \omega_a{}^b - \omega_a{}^c \wedge \omega_c{}^b \\ &= \mathrm{D}\, \omega_a{}^b - \omega_c{}^b \wedge \omega_a{}^c. \end{aligned} \tag{8}$$

The 3$^{rd}$ ordinary exterior differentials of $\mathsf{e}_a$ are given by

$$\begin{aligned} \mathrm{d}^3\, \mathsf{e}_a &= \mathrm{d}\, \mathrm{d}^2\, \mathsf{e}_a \\ &= \mathrm{d}\, \mathsf{e}_b\, \Omega_a{}^b \\ &= (\mathrm{d}\, \mathsf{e}_b) \wedge \Omega_a{}^b + \mathsf{e}_b\, \mathrm{d}\, \Omega_a{}^b \\ &= (\mathsf{e}_c\, \omega_b{}^c) \wedge \Omega_a{}^b + \mathsf{e}_b\, \mathrm{d}\, \Omega_a{}^b \\ &= \mathsf{e}_c\, \Omega_b{}^c \wedge \omega_a{}^b \end{aligned} \tag{9}$$

using Bianchi's identity for $\Omega_a{}^b$,[17] i.e.,

$$\begin{aligned} \mathrm{D}\, \Omega_a{}^b &= \mathrm{d}\, \Omega_a{}^b - \omega_a{}^c \wedge \Omega_c{}^b + \omega_c{}^b \wedge \Omega_a{}^c \\ &= 0, \end{aligned} \tag{10}$$

as well as by

$$\begin{aligned} \mathrm{d}^3\, \mathsf{e}_a &= \mathrm{d}^2\, \mathrm{d}\, \mathsf{e}_a \\ &= \mathrm{d}^2\, \mathsf{e}_b\, \omega_a{}^b \\ &= (\mathrm{d}^2\, \mathsf{e}_b) \wedge \omega_a{}^b + \mathsf{e}_b\, \mathrm{d}^2\, \omega_a{}^b \\ &= (\mathsf{e}_c\, \Omega_b{}^c) \wedge \omega_a{}^b + 0 \\ &= \mathsf{e}_c\, \Omega_b{}^c \wedge \omega_a{}^b \end{aligned} \tag{11}$$

1 Eguchi, T., P. B. Gilkey, and A. J. Hanson, "Gravitation, Gauge Theories and Differential Geometry," *Physics Reports*, **66** (1980) 213.
2 Nakahara, M., *Geometry, Topology, and Physics*, IOP Publishing Ltd., Bristol, England (1990), p. 385.
3 Bradlow, S., "Characteristic Classes for Complex Bundles II," in *The Geometry of Vector Bundles and an Introduction to Gauge Theory*, [unpublished lecture notes], 1998, Lecture 30.
4 Misner, C. W., K. S. Thorne, and J. A. Wheeler, *Gravitation*, W. H. Freeman & Co., San Francisco, CA (1973), p. 350.
5 Cartan, É. J., *Leçons sur la géométrie des espaces de Riemann*, Gauthier-Villars et C$^{ie}$, Paris, France (1928), p. 180; *Leçons sur la géométrie des espaces de Riemann, 2nd édition, rev. et augm.*, Gauthier-Villars, Paris, France (1946), p. 180.
6 Chern, S. S., "A simple intrinsic proof of the Gauss-Bonnet formula for closed Riemannian manifolds," *Ann. of Math.*, **45** (1944) 747; "On the curvatura integra in a Riemannian manifold," *ibid.*, **46** (1945) 674; "On the Minimal Immersions of the Two-sphere in a Space of Constant Curvature," in Gunning, R. C. (ed.), *Problems in Analysis: A Symposium in Honor of Salomon Bochner*, Princeton University Press, Princeton, NJ (1970), p. 27.
7 Bott, R., and S. S. Chern, "Some Formulas Related to Complex Transgression," in Haefliger, A., and R. Narasimhan (eds.), *Essays on Topology and Related Topics; Memoires dédiés à Georges de Rham*, Springer-Verlag, New York, NY (1970), p. 48.
8 Bradlow, S., "Curvatures," *op. cit.*, Lecture 20.
9 Cartan, É. J., "Sur les variétés à connexion affine et la théorie de la relativité généralisée (Première Partie)," *Annales scientifiques de l'École Normale supérieure*, **40** (1923) 325; (1928), *op. cit.*, p. 182; (1946), *op. cit.*, p. 181.
10 Chern, S. S., "Differential Geometry of Fiber Bundles," in *Proceedings of the International Congress of Mathematicians*, **2** (1950) 397; "Topics in Differential Geometry," [mimeographed notes], The Institute for Advanced Study, Princeton, NJ (1951), Chap. 3, Sec. 3; (1970), *op. cit.*
11 Cohen, J. M., *Exterior Differential Forms and Orthonormal Cartan Frames*, [unpublished monograph], (1975), p. 33.
12 Flanders, H., "Development of an Extended Exterior Differential Calculus," *Trans. Amer. Math. Soc.*, **75** (1953) 311; "Methods in Affine Connection Theory," *Pac. J. Math.*, **5** (1955) 391; *Differential Forms with Applications to the Physical Sciences*, Academic Press, New York, NY (1963), pp. 130 and 144.
13 Misner, C. W., *et al.*, *op. cit.*, pp. 350-351 and 359.
14 Petrie, T., and J. Randall, *Connections, Definite Forms, and Four-manifolds*, Clarendon Press, Oxford, England (1990), pp. 11-12.
15 Sattinger, D. H., and O. L. Weaver, *Lie Groups and Algebras with Applications to Physics, Geometry, and Mechanics*, Springer-Verlag, New York, NY (1986), p. 102.
16 Trautmann, A., *Differential Geometry for Physicists—Stony Brook Lectures*, Bibliopolis, edizioni di filosofia e scienze, Napoli, Italy (1984), pp. 91-92.
17 Cartan, É. J., (1928), *op. cit.*, p. 214; (1946), *op. cit.*, p. 210.

using Poincaré's theorem for scalar-valued exterior differential forms,[18-19] i.e.,

$$\mathrm{d}^2\,\alpha = 0, \tag{12}$$

where $\alpha$ is an arbitrary scalar-valued exterior differential form.

The $4^{th}$ ordinary exterior differentials of $\mathrm{e}_a$ are given by

$$\begin{aligned}\mathrm{d}^4\,\mathrm{e}_a &= \mathrm{d}^2\,\mathrm{d}^2\,\mathrm{e}_a \\ &= \mathrm{d}^2\,\mathrm{e}_b\,\Omega_a{}^b \\ &= (\mathrm{d}^2\,\mathrm{e}_b)\wedge\Omega_a{}^b + \mathrm{e}_b\,\mathrm{d}^2\,\Omega_a{}^b \\ &= (\mathrm{e}_c\,\Omega_b{}^c)\wedge\Omega_a{}^b + 0 \\ &= \mathrm{e}_c\,\Omega_b{}^c\wedge\Omega_a{}^b.\end{aligned} \tag{13}$$

The $p^{th}$ ordinary exterior differentials of $\mathrm{e}_a$ for $0 \le p \le 10$ are given by

$$\mathrm{e}_a = \mathrm{e}_b\,\delta_a^b, \tag{14}$$

$$\mathrm{d}\,\mathrm{e}_a = \mathrm{e}_b\,\omega_a{}^b, \tag{15}$$

$$\mathrm{d}^2\,\mathrm{e}_a = \mathrm{e}_b\,\Omega_a{}^b, \tag{16}$$

$$\mathrm{d}^3\,\mathrm{e}_a = \mathrm{e}_c\,\Omega_b{}^c\wedge\omega_a{}^b, \tag{17}$$

$$\mathrm{d}^4\,\mathrm{e}_a = \mathrm{e}_c\,\Omega_b{}^c\wedge\Omega_a{}^b, \tag{18}$$

$$\mathrm{d}^5\,\mathrm{e}_a = \mathrm{e}_d\,\Omega_c{}^d\wedge\Omega_b{}^c\wedge\omega_a{}^b, \tag{19}$$

$$\mathrm{d}^6\,\mathrm{e}_a = \mathrm{e}_d\,\Omega_c{}^d\wedge\Omega_b{}^c\wedge\Omega_a{}^b, \tag{20}$$

$$\mathrm{d}^7\,\mathrm{e}_a = \mathrm{e}_e\,\Omega_d{}^e\wedge\Omega_c{}^d\wedge\Omega_b{}^c\wedge\omega_a{}^b, \tag{21}$$

$$\mathrm{d}^8\,\mathrm{e}_a = \mathrm{e}_e\,\Omega_d{}^e\wedge\Omega_c{}^d\wedge\Omega_b{}^c\wedge\Omega_a{}^b, \tag{22}$$

$$\mathrm{d}^9\,\mathrm{e}_a = \mathrm{e}_f\,\Omega_e{}^f\wedge\Omega_d{}^e\wedge\Omega_c{}^d\wedge\Omega_b{}^c\wedge\omega_a{}^b, \tag{23}$$

$$\mathrm{d}^{10}\,\mathrm{e}_a = \mathrm{e}_f\,\Omega_e{}^f\wedge\Omega_d{}^e\wedge\Omega_c{}^d\wedge\Omega_b{}^c\wedge\Omega_a{}^b. \tag{24}$$

In general, the $p^{th}$ ordinary exterior differentials of $\mathrm{e}_a$ for $p > 0$ are given (cf. Flanders[20]) by

$$\begin{aligned}\mathrm{d}^p\,\mathrm{e}_a &= \begin{cases}\mathrm{e}_{i_1}\,\Omega_{i_2}{}^{i_1}\wedge\Omega_{i_3}{}^{i_2}\wedge\Omega_{i_4}{}^{i_3}\wedge\ldots\wedge\Omega_{i_{(p-1)/2}}{}^{i_{(p-3)/2}}\wedge\Omega_{i_{(p+1)/2}}{}^{i_{(p-1)/2}}\wedge\omega_a{}^{i_{(p+1)/2}}, & \text{if } p \text{ is odd}\\ \mathrm{e}_{i_1}\,\Omega_{i_2}{}^{i_1}\wedge\Omega_{i_3}{}^{i_2}\wedge\Omega_{i_4}{}^{i_3}\wedge\ldots\wedge\Omega_{i_{(p-2)/2}}{}^{i_{(p-4)/2}}\wedge\Omega_{i_{p/2}}{}^{i_{(p-2)/2}}\wedge\Omega_a{}^{i_{p/2}}, & \text{if } p \text{ is even}\end{cases}\\ &= \begin{cases}\mathrm{e}_{i_1}\,\omega_a{}^{i_{(p+1)/2}}\wedge\Omega_{i_{(p+1)/2}}{}^{i_{(p-1)/2}}\wedge\Omega_{i_{(p-1)/2}}{}^{i_{(p-3)/2}}\wedge\ldots\wedge\Omega_{i_4}{}^{i_3}\wedge\Omega_{i_3}{}^{i_2}\wedge\Omega_{i_2}{}^{i_1}, & \text{if } p \text{ is odd}\\ \mathrm{e}_{i_1}\,\Omega_a{}^{i_{p/2}}\wedge\Omega_{i_{p/2}}{}^{i_{(p-2)/2}}\wedge\Omega_{i_{(p-2)/2}}{}^{i_{(p-4)/2}}\wedge\ldots\wedge\Omega_{i_4}{}^{i_3}\wedge\Omega_{i_3}{}^{i_2}\wedge\Omega_{i_2}{}^{i_1}, & \text{if } p \text{ is even}\end{cases}\\ &= \begin{cases}\mathrm{e}_{i_{(p+1)/2}}\,\omega_a{}^{i_1}\wedge\Omega_{i_1}{}^{i_2}\wedge\Omega_{i_2}{}^{i_3}\wedge\ldots\wedge\Omega_{i_{(p-5)/2}}{}^{i_{(p-3)/2}}\wedge\Omega_{i_{(p-3)/2}}{}^{i_{(p-1)/2}}\wedge\Omega_{i_{(p-1)/2}}{}^{i_{(p+1)/2}}, & \text{if } p \text{ is odd}\\ \mathrm{e}_{i_{p/2}}\,\Omega_a{}^{i_1}\wedge\Omega_{i_1}{}^{i_2}\wedge\Omega_{i_2}{}^{i_3}\wedge\ldots\wedge\Omega_{i_{(p-6)/2}}{}^{i_{(p-4)/2}}\wedge\Omega_{i_{(p-4)/2}}{}^{i_{(p-2)/2}}\wedge\Omega_{i_{(p-2)/2}}{}^{i_{p/2}}, & \text{if } p \text{ is even}\end{cases},\end{aligned} \tag{25}$$

the contractions of which with $\omega^a$ are given by

$$\begin{aligned}\langle\omega^a, \mathrm{d}^p\,\mathrm{e}_a\rangle &= \begin{cases}\delta_{i_{(p+1)/2}}^{\;a}\,\omega_a{}^{i_1}\wedge\Omega_{i_1}{}^{i_2}\wedge\Omega_{i_2}{}^{i_3}\wedge\ldots\wedge\Omega_{i_{(p-5)/2}}{}^{i_{(p-3)/2}}\wedge\Omega_{i_{(p-3)/2}}{}^{i_{(p-1)/2}}\wedge\Omega_{i_{(p-1)/2}}{}^{i_{(p+1)/2}}, & \text{if } p \text{ is odd}\\ \delta_{i_{p/2}}^{a}\,\Omega_a{}^{i_1}\wedge\Omega_{i_1}{}^{i_2}\wedge\Omega_{i_2}{}^{i_3}\wedge\ldots\wedge\Omega_{i_{(p-6)/2}}{}^{i_{(p-4)/2}}\wedge\Omega_{i_{(p-4)/2}}{}^{i_{(p-2)/2}}\wedge\Omega_{i_{(p-2)/2}}{}^{i_{p/2}}, & \text{if } p \text{ is even}\end{cases}\\ &= \begin{cases}\omega_{i_{(p+1)/2}}{}^{i_1}\wedge\Omega_{i_1}{}^{i_2}\wedge\Omega_{i_2}{}^{i_3}\wedge\ldots\wedge\Omega_{i_{(p-5)/2}}{}^{i_{(p-3)/2}}\wedge\Omega_{i_{(p-3)/2}}{}^{i_{(p-1)/2}}\wedge\Omega_{i_{(p-1)/2}}{}^{i_{(p+1)/2}}, & \text{if } p \text{ is odd}\\ \Omega_{i_{p/2}}{}^{i_1}\wedge\Omega_{i_1}{}^{i_2}\wedge\Omega_{i_2}{}^{i_3}\wedge\ldots\wedge\Omega_{i_{(p-6)/2}}{}^{i_{(p-4)/2}}\wedge\Omega_{i_{(p-4)/2}}{}^{i_{(p-2)/2}}\wedge\Omega_{i_{(p-2)/2}}{}^{i_{p/2}}, & \text{if } p \text{ is even}\end{cases}.\end{aligned} \tag{26}$$

Equation (26) then yields the formula in question, viz.,

$$ch_{(p)} = \frac{i^p}{2^p\pi^p p!}\,\langle\omega^a, \mathrm{d}^{2p}\,\mathrm{e}_a\rangle. \tag{27}$$

For an aside, note that using Eq. (18) and taking the $0^{th}$ ordinary exterior differentials of $\mathrm{e}_a$ to be given by

$$\mathrm{d}^0\,\mathrm{e}_a = \mathrm{e}_a, \tag{28}$$

the polynomial $ch_{(p)}$ for $p = 0$ is given by

$$\begin{aligned}ch_{(0)} &= \frac{i^0}{2^0\pi^0 0!}\,\langle\omega^a, \mathrm{d}^0\,\mathrm{e}_a\rangle \\ &= \langle\omega^a, \mathrm{e}_a\rangle \\ &= \delta_a^a \\ &= n,\end{aligned} \tag{29}$$

where $n$, having appeared above, is the number of dimensions of $M$ (cf. Bradlow[21]).

Expressions for Chern characters and their formulas for $0 \le p \le 15$ appear in Table 1 (see below).

---

[18] Cartan, É. J., (1928), *op. cit.*, pp. 209 and 215; *Le systèmes différentiels extérieurs et leurs applications géometriques*, Actualités scientifiques et industrielles, 994, Exposés de géométrie, vol. 12, Librairie scientifiques Hermann & C$^{ie}$, Éditeurs, Paris, France (1945), p. 37; (1946), *op. cit.*, pp. 208 and 211.

[19] Chern, S. S., "The Geometry of Isotropic Surfaces," *Annals of Math.*, **43** (1942) 545; "On the Euclidean Connections in a Finsler Space," *Proc. Nat. Acad. Sci.*, **29** (1943) 33; "Local Equivalence and Euclidean Connections in Finsler Spaces," *Science Reports Nat. Tsing Hua Univ.*, **5** (1948) 95.

[20] Flanders, H., (1953), *op. cit.*, Eq. (9.5).

[21] Bradlow, S., "Characteristic Classes for Complex Bundles II," *op. cit.*

TABLE 1. EXPRESSIONS FOR CHERN CHARACTERS AND THEIR FORMULAS FOR $0 \le p \le 15$

| ORDER | CURVATURE DEPENDENCE | QUANTITY | MINIMUM DIMENSIONALITY OF $M$ | FORMULA | $p^{th}$ CHERN CHARACTER |
|---|---|---|---|---|---|
| $p$ | — | $ch_{(p)}$ | $2p$ | $\frac{i^p}{2^p \pi^p p!} \langle \omega^a, d^{2p} e_a \rangle$ | $\frac{i^p}{2^p \pi^p p!} \mathrm{tr}(\Omega^p)$ |
| 0 | Zero | $ch_{(0)}$ | 0 | $\frac{i^0}{2^0 \pi^0 0!} \langle \omega^a, d^0 e_a \rangle$ | $n$ |
| 1 | Linear | $ch_{(1)}$ | 2 | $\frac{i^1}{2^1 \pi^1 1!} \langle \omega^a, d^2 e_a \rangle$ | $\frac{i}{2\pi} \mathrm{tr}(\Omega)$ |
| 2 | Quadratic | $ch_{(2)}$ | 4 | $\frac{i^2}{2^2 \pi^2 2!} \langle \omega^a, d^4 e_a \rangle$ | $\frac{-1}{8\pi^2} \mathrm{tr}(\Omega^2)$ |
| 3 | Cubic | $ch_{(3)}$ | 6 | $\frac{i^3}{2^3 \pi^3 3!} \langle \omega^a, d^6 e_a \rangle$ | $\frac{-i}{48\pi^3} \mathrm{tr}(\Omega^3)$ |
| 4 | Quartic | $ch_{(4)}$ | 8 | $\frac{i^4}{2^4 \pi^4 4!} \langle \omega^a, d^8 e_a \rangle$ | $\frac{1}{384\pi^4} \mathrm{tr}(\Omega^4)$ |
| 5 | Quintic | $ch_{(5)}$ | 10 | $\frac{i^5}{2^5 \pi^5 5!} \langle \omega^a, d^{10} e_a \rangle$ | $\frac{i}{3840\pi^5} \mathrm{tr}(\Omega^5)$ |
| 6 | Sextic | $ch_{(6)}$ | 12 | $\frac{i^6}{2^6 \pi^6 6!} \langle \omega^a, d^{12} e_a \rangle$ | $\frac{-1}{46{,}080\pi^6} \mathrm{tr}(\Omega^6)$ |
| 7 | Septic | $ch_{(7)}$ | 14 | $\frac{i^7}{2^7 \pi^7 7!} \langle \omega^a, d^{14} e_a \rangle$ | $\frac{-i}{645{,}120\pi^7} \mathrm{tr}(\Omega^7)$ |
| 8 | Octic | $ch_{(8)}$ | 16 | $\frac{i^8}{2^8 \pi^8 8!} \langle \omega^a, d^{16} e_a \rangle$ | $\frac{1}{10{,}321{,}920\pi^8} \mathrm{tr}(\Omega^8)$ |
| 9 | Nonic | $ch_{(9)}$ | 18 | $\frac{i^9}{2^9 \pi^9 9!} \langle \omega^a, d^{18} e_a \rangle$ | $\frac{i}{185{,}794{,}560\pi^9} \mathrm{tr}(\Omega^9)$ |
| 10 | Decic | $ch_{(10)}$ | 20 | $\frac{i^{10}}{2^{10} \pi^{10} 10!} \langle \omega^a, d^{20} e_a \rangle$ | $\frac{-1}{3{,}715{,}891{,}200\pi^{10}} \mathrm{tr}(\Omega^{10})$ |
| 11 | Undecic | $ch_{(11)}$ | 22 | $\frac{i^{11}}{2^{11} \pi^{11} 11!} \langle \omega^a, d^{22} e_a \rangle$ | $\frac{-i}{81{,}749{,}606{,}400\pi^{11}} \mathrm{tr}(\Omega^{11})$ |
| 12 | Duodecic | $ch_{(12)}$ | 24 | $\frac{i^{12}}{2^{12} \pi^{12} 12!} \langle \omega^a, d^{24} e_a \rangle$ | $\frac{1}{1{,}961{,}990{,}553{,}600\pi^{12}} \mathrm{tr}(\Omega^{12})$ |
| 13 | Tredecic | $ch_{(13)}$ | 26 | $\frac{i^{13}}{2^{13} \pi^{13} 13!} \langle \omega^a, d^{26} e_a \rangle$ | $\frac{i}{51{,}011{,}754{,}393{,}600\pi^{13}} \mathrm{tr}(\Omega^{13})$ |
| 14 | Quattuordecic | $ch_{(14)}$ | 28 | $\frac{i^{14}}{2^{14} \pi^{14} 14!} \langle \omega^a, d^{28} e_a \rangle$ | $\frac{-1}{1{,}428{,}329{,}123{,}020{,}800\pi^{14}} \mathrm{tr}(\Omega^{14})$ |
| 15 | Quindecic | $ch_{(15)}$ | 30 | $\frac{i^{15}}{2^{15} \pi^{15} 15!} \langle \omega^a, d^{30} e_a \rangle$ | $\frac{-i}{42{,}849{,}873{,}690{,}624{,}000\pi^{15}} \mathrm{tr}(\Omega^{15})$ |
| 16 | Sexdecic | $ch_{(16)}$ | 32 | $\frac{i^{16}}{2^{16} \pi^{16} 16!} \langle \omega^a, d^{32} e_a \rangle$ | $\frac{1}{1{,}371{,}195{,}958{,}099{,}968{,}000\pi^{16}} \mathrm{tr}(\Omega^{16})$ |
| 17 | Septendecic | $ch_{(17)}$ | 34 | $\frac{i^{17}}{2^{17} \pi^{17} 17!} \langle \omega^a, d^{34} e_a \rangle$ | $\frac{i}{46{,}620{,}662{,}575{,}398{,}912{,}000\pi^{17}} \mathrm{tr}(\Omega^{17})$ |
| 18 | Octodecic | $ch_{(18)}$ | 36 | $\frac{i^{18}}{2^{18} \pi^{18} 18!} \langle \omega^a, d^{36} e_a \rangle$ | $\frac{-1}{1{,}678{,}343{,}852{,}714{,}360{,}832{,}000\pi^{18}} \mathrm{tr}(\Omega^{18})$ |
| 19 | Novemdecic | $ch_{(19)}$ | 38 | $\frac{i^{19}}{2^{19} \pi^{19} 19!} \langle \omega^a, d^{38} e_a \rangle$ | $\frac{-i}{63{,}777{,}066{,}403{,}145{,}711{,}616{,}000\pi^{19}} \mathrm{tr}(\Omega^{19})$ |
| 20 | Vigintic | $ch_{(20)}$ | 40 | $\frac{i^{20}}{2^{20} \pi^{20} 20!} \langle \omega^a, d^{40} e_a \rangle$ | $\frac{1}{2{,}551{,}082{,}656{,}125{,}828{,}464{,}640{,}000\pi^{20}} \mathrm{tr}(\Omega^{20})$ |